# Assessing correlations of perovskite catalytic performance with electronic structure descriptors

**Authors:** Ryan Jacobs[1], Jonathan Hwang[2], Yang Shao-Horn[2], Dane Morgan[1]

[1] Department of Materials Science and Engineering, University of Wisconsin-Madison, Madison, WI, 53706, USA

[2] Department of Materials Science and Engineering, Massachusetts Institute of Technology, Cambridge, MA, 02139, USA

## Abstract

Electronic structure descriptors are computationally efficient quantities used to construct qualitative correlations for a variety of properties. In particular, the oxygen $p$-band center has been used to guide material discovery and fundamental understanding of an array of perovskite compounds for use in catalyzing the oxygen reduction and evolution reactions. However, an assessment of the effectiveness of the oxygen $p$-band center at predicting key measures of perovskite catalytic activity has not been made, and would be highly beneficial to guide future predictions and codify best practices. Here, we have used Density Functional Theory at the PBE, PBEsol, PBE+$U$, SCAN and HSE06 levels to assess the correlations of numerous measures of catalytic performance for a series of technologically relevant perovskite oxides, using the bulk oxygen $p$-band center as an electronic structure descriptor. We have analyzed correlations of the calculated oxygen $p$-band center for all considered functionals with the experimentally measured X-ray emission spectroscopy oxygen $p$-band center and multiple measures of catalytic activity, including high temperature oxygen reduction surface exchange rates, aqueous oxygen evolution current densities, and binding energies of oxygen evolution intermediate species. Our results show that the best correlations for all measures of catalytic activity considered here are made with PBE-level calculations, with strong observed linear correlations with the bulk oxygen $p$-band center ($R^2$ = 0.81-0.87). This study shows that strong linear correlations between numerous important measures of catalytic activity and the oxygen $p$-band bulk descriptor can be obtained under a consistent computational framework, and these correlations can serve as a guide for future experiments and simulations for development of perovskite and related oxide catalysts.

# Table of Contents Figure

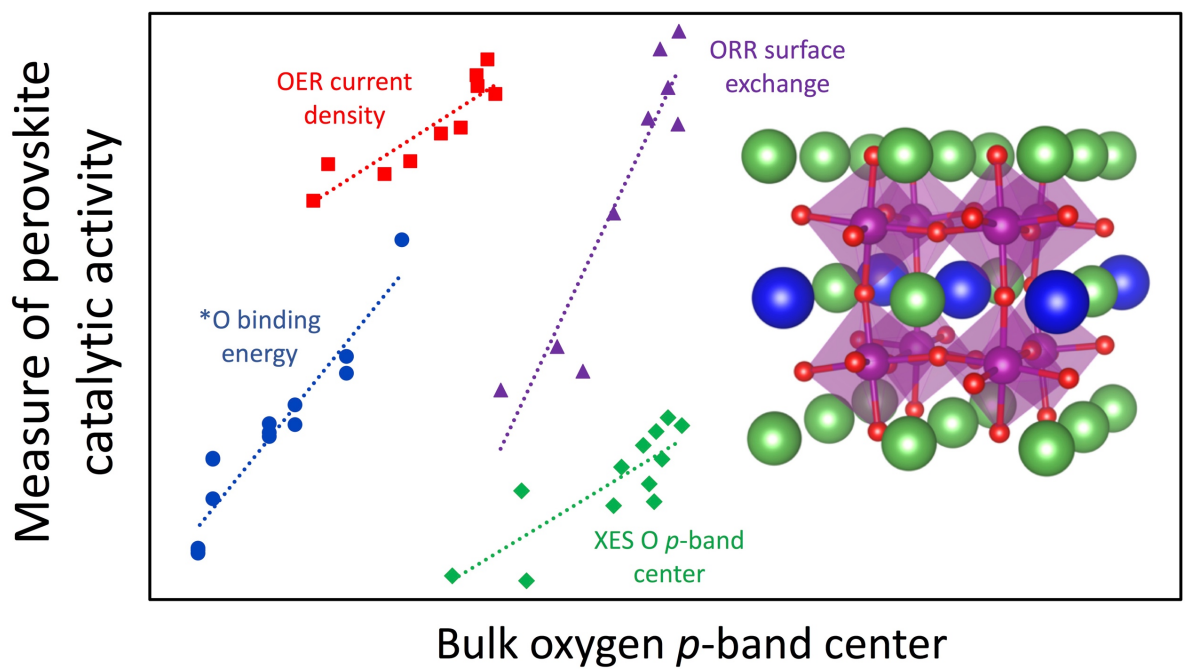

# Main

## 1. Introduction

Descriptors computed from first-principles methods have been successfully used to explain and predict the catalytic activity of precious metals, metal alloys, and oxides for a number of chemical reactions over the past 25 years.[1–4] The utility of descriptors arises from their ability to provide fast, qualitative predictions and fundamental understanding of complex physical processes, such as the catalytic activity of chemical reactions on the surfaces of materials. First-principles calculated descriptors also enable high-throughput computational materials design of new or improved materials by serving as a rapid materials screening tool.[5] The pioneering work of Hammer and Nørskov involved the development of the metal $d$-band center (defined as the centroid of the projected density of states of the metal $d$ electrons relative to the Fermi level) applied as an electronic structure descriptor to predict the catalytic activity of precious metal and metal alloy catalysts to hydrogen[1–3,5–7] and oxygen evolution[3,6,7] and oxygen reduction[3,6–9] for electrochemical water splitting.[1–4,10] This descriptor has been instrumental in facilitating the computational design of new metal alloy catalysts and has enabled improved fundamental insight of the surface reactivity of metals.[5–9]

More recently, the use of descriptors has also been applied to perovskite oxide (chemical formula $A_{1-x}A'_xB_{1-y}B'_yO_{3-\delta}$, see **Figure 1**) catalysts. The O *p*-band center has been used as a bulk electronic structure descriptor to predict numerous properties of perovskite (and the closely related Ruddlesden-Popper) oxides. The O *p*-band center is defined as the centroid of the projected density of states of the oxygen 2*p* orbitals relative to the Fermi level, as shown schematically in **Figure 1**, and information regarding the band center calculations can be found in **Section 4**. Properties predicted using the O *p*-band center include: vacancy and interstitial formation and migration energies,[11–14] work functions,[15] and catalytic properties such as the high temperature surface exchange rate for the oxygen reduction reaction (ORR) in solid oxide fuel cells,[11,16,17] binding energies of reaction intermediates for aqueous oxygen reduction and evolution reactions,[18,19] and overpotentials of oxygen evolution reaction (OER) and ORR in basic solution.[17,20] However, discrepancies exist among predictions made using different Density Functional Theory (DFT) exchange and correlation functionals, and from experiments. For example, there have been reported discrepancies between DFT calculations using different functionals and experimental values of OER and ORR potentials and trends of intermediate binding energies.[19,21–23] Understanding the origin of these discrepancies and providing recommendations of best practices when employing the O *p*-band center as a descriptor is the main purpose of this work. The results and understanding developed in this work will help ensure the most accurate computational predictions of an array of technologically relevant perovskite properties, provide recommendations of best practices when using this descriptor to promote consistency and transferability of results in the literature, and help guide future simulations and experiments related to revealing trends in perovskite catalytic activity.

In this work, we have used DFT to calculate the electronic structure properties of 20 perovskite oxides using five different DFT exchange and correlation functionals: PBE,[24] PBEsol,[25] PBE+*U*,[26–28] SCAN,[29,30] and HSE06 (henceforth "HSE").[31,32] The same *U* values in PBE+*U* as those used in the Materials Project were used: $U$ = 3.25, 3.7, 3.9, 5.3, 3.32, and 6.7 eV for V, Cr, Mn, Fe, Co, and Ni, respectively.[33] HSE calculations were performed using a cutoff of 0.2 Å$^{-1}$ and Hartree-Fock exchange fraction of 25% (standard HSE06). The structure of the perovskites studied in this work is shown in **Figure 1**. We chose these particular 20 perovskite compounds to represent a wide range of chemistries by sampling the entire 3*d* row of transition metals. The chosen materials are all technologically relevant, with an emphasis on materials for fast oxygen surface exchange and transport.[21,34–37] These materials are, in order of their 3$d$ electron count calculated using formal valences: LaScO$_3$, SrTiO$_3$, LaTiO$_3$, SrVO$_3$, LaVO$_3$, LaCrO$_3$, La$_{0.75}$Sr$_{0.25}$MnO$_3$ (LSM25), LaMnO$_3$, SrFeO$_{2.75}$, Ba$_{0.5}$Sr$_{0.5}$Co$_{0.75}$Fe$_{0.25}$O$_{2.625}$ (BSCF2.625), LaFeO$_3$, SrCoO$_3$, GdBaCo$_2$O$_{5.5}$ (GBCO, double perovskite layered ordering), La$_{0.5}$Sr$_{0.5}$CoO$_3$ (LSC50), PrBaCo$_2$O$_{5.5}$ (PBCO, double perovskite layered ordering), SmBaCo$_2$O$_{5.5}$ (SBCO, double perovskite layered ordering), La$_{0.75}$Sr$_{0.25}$CoO$_3$ (LSC25), LaCoO$_3$, LaNiO$_3$, and LaCuO$_3$. Additional calculation details can be found in **Section 4**.

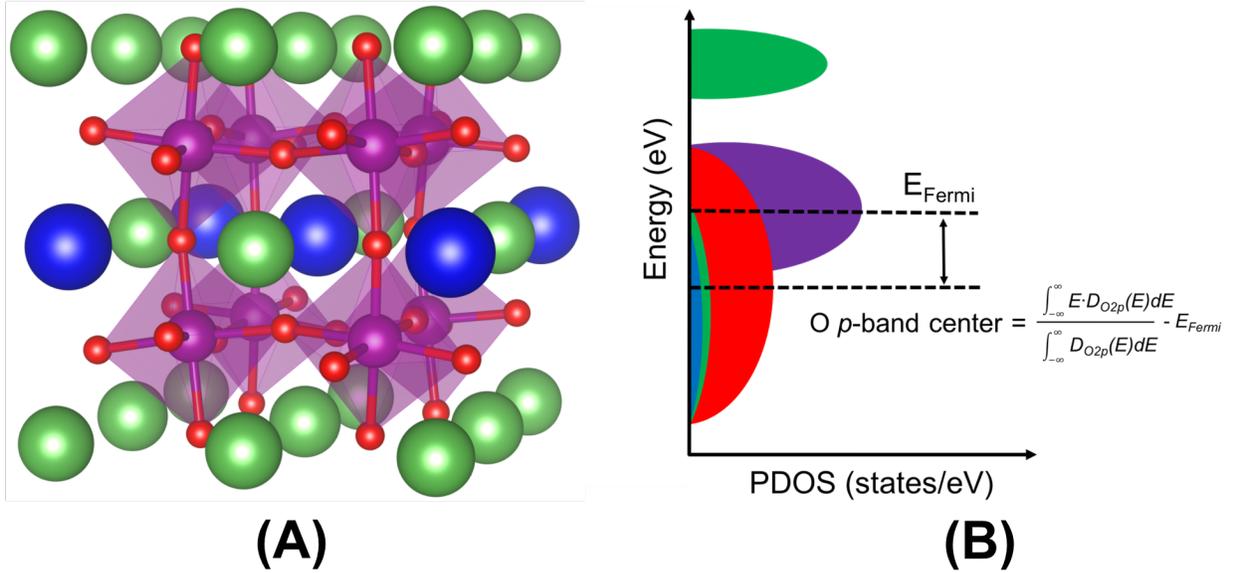

**Figure 1.** (A) Representative pseudocubic perovskite structure used in this study. This example is for La$_{0.75}$Sr$_{0.25}$MnO$_3$ (LSM25), where the green, blue, purple, and red colored spheres denote La, Sr, Mn, and O, respectively. (B) Schematic projected density of states (PDOS) plot demonstrating the definition of the O $p$-band center descriptor as the centroid of the projected oxygen 2$p$ orbitals, relative to the Fermi level. The green, blue, purple and red colors represent the La, Sr, Mn, and O PDOS near the Fermi level, respectively. The Fermi level and O $p$-band center are marked with dashed black lines.

We have specifically investigated five aspects of O $p$-band descriptor behavior. (1) We studied trends in the O $p$-band center and transition metal (TM) $d$-band center as a function of composition and DFT functional (**Section 2.1**). Examination of these trends is valuable to understand how the electronic structure of these materials evolves as different DFT modeling methods are employed, and how this evolution couples to the material composition. (2) The ability of DFT to predict experimental O $p$-band center as obtained using X-ray emission spectroscopy (XES), and the overall functional relationship between calculated and experimental O $p$-band center values (**Section 2.2**). This comparison to experiment is useful as XES is used to obtain quantitative experimental information on the element-specific electronic structure near the Fermi level, providing key insights into the catalytic activity of materials through the analysis of the energy levels in a material relative to well-defined redox energy levels.[38,39] (3) The ability of DFT to predict experimental values of the high temperature surface exchange coefficient $k^*$ (**Section 2.3**). This offers valuable insight as $k^*$ is related to the overall catalytic rate of the oxygen reduction reaction (ORR), a key chemical reaction in devices utilizing the transport and exchange of oxygen with the environment, such as solid oxide fuel cells (SOFCs), gas separation membranes, and chemical looping applications,[11,16,40–43] (4) the ability of DFT to predict experimental aqueous OER current densities (**Section 2.4**) and (5) correlations between OER intermediate species binding energies and the O *p*-band center (**Section 2.5**). Effectively catalyzing the OER is necessary to produce molecular $H_2$ and $O_2$ from water for use as chemical fuels. The current density produced at a constant applied voltage provides a measure of the OER catalytic activity, with a higher current density (at constant voltage) indicative of a more active catalyst.[17,20,44,45]

## 2. Results and Discussion

**2.1: Chemical trends in oxygen and transition metal band centers**

In this section, we have developed an understanding of the trends in the O *p*- and transition metal (TM) *d*-band center descriptors as a function of perovskite composition and DFT functional employed. To do this, we have examined the chemical trends in the O *p*-band center, the TM *d*-band center, and the difference between the O *p*- and TM *d*-band center (referred to as (O *p* – TM *d*) band center) for all materials and DFT functionals used in this study. In addition, we have assessed shifts in the (O *p* – TM *d*) band center for each functional relative to PBE. Analysis of these chemical trends is useful as it allows for straightforward examination of the changes in bulk electronic structure resulting from changes in perovskite chemistry as well as changes in the electronic structure arising from the use of different DFT functionals, and connection of these physical changes to the known catalytic trends of these materials to establish catalyst design principles.

By examining the trends of O $p$-band center in **Figure 2A**, the oxides with the highest (most positive) O $p$-band center are either insulating ($3d^0$, LaScO$_3$ and SrTiO$_3$), have a full $3d$ band ($3d^8$, LaCuO$_3$), or have $3d$ electron counts between about 4-5.5 electrons (for example, BSCF2.625 at $3d^{4.5}$ and PBCO at $3d^{5.5}$). Materials such as BSCF2.625 and PBCO have some of the highest O $p$-band center values due to the large degree of O $p$ – TM $d$-band hybridization, which stems from the low degree of bonding ionicity in these compounds. Such large band hybridization results in a Fermi level comprised of highly overlapping TM $3d$ and O $2p$ states and a high value of the O $p$-band center. Conversely, the materials with the lowest (most negative) O $p$-band center have either 1 or 2 $3d$ electrons ($3d^1$-$3d^2$, LaTiO$_3$, SrVO$_3$, LaVO$_3$). The reason for these low $3d$ electron count materials having low O $p$-band center values is the high bond ionicity and high $d$-band center energy relative to the O $p$-band center, resulting in very little O $p$ – TM $d$ bond hybridization and a corresponding low O $p$-band center. It is worth noting that insulating materials such as LaScO$_3$ and SrTiO$_3$ and full $3d$ band materials such as LaCuO$_3$ also have predominantly ionic bonding and have some of the highest O $p$-band center values of the materials examined here. The high O $p$-band center values for these materials arises from the $3d$ band being either empty (LaScO$_3$ and SrTiO$_3$) or completely full (LaCuO$_3$), resulting in a Fermi level that is comprised almost entirely of O $2p$ states and a high value of the O $p$-band center. These trends in O $p$-band center with $3d$ electron count are also consistent with trends in calculated perovskite work functions with O $p$-band center and $3d$ number as reported in Jacobs et al.[15,46] Based on the trends in TM $d$-band center in **Figure 2B**, qualitatively, for all functionals, when progressing from low to high $3d$ electron count, the $3d$ band tends to shift down in energy, with a higher fraction of the $3d$ band being filled. This trend of lower TM $d$-band center with higher $3d$ electron count is qualitatively consistent with previous studies of $d$-band center and number of $d$ electrons for metals.[1–3] Moreover, this trend is consistent with redox energies in lithium ion battery cathode materials, where higher cathode voltages (redox energies versus the Li/Li$^+$ redox couple) are observed for positive electrode materials containing the late transition metals Fe, Co and Ni compared to early transition metals such as Ti and V.[47–51] The HSE data in **Figure 2B** clearly shows the highest TM $d$-band centers compared to the other functionals for low $3d$ electron numbers, which is consistent with HSE creating larger bandgaps (compared to the other functionals investigated here) in these band-insulating and Mott-Hubbard insulating oxides, with a large upshift of the unoccupied $3d$ states relative to the Fermi level.[30,32,47,52]

The value of (O $p$ – TM $d$) band center tends to increase with increasing 3$d$ electron number, as shown in **Figure 2C**. The value of the (O $p$ – TM $d$) band center is expected to be qualitatively correlated with the overlap of the O 2$p$ and TM 3$d$ bands, which is a measure of the bond hybridization (i.e. covalency) of the O and TM species.[38,39,53] Therefore, this trend of increasing (O $p$ – TM $d$) band center with 3$d$ electron number can be understood as arising from the TM-O bond covalency increasing as 3$d$ electron number increases. This observed trend of increasing covalency with 3$d$ electron number is fully consistent with previous DFT and experimental X-ray absorption spectroscopy studies.[15,38,39] A value of (O $p$ – TM $d$) band center far from zero (little band overlap) indicates a less covalent (more ionic) compound while a value of (O $p$ – TM $d$) band center near zero indicates a highly covalent compound. In **Figure 2C**, all functionals show the same general trend of increasing covalency up to approximately 3$d^{4.5}$ and nearly no change in covalency for 3$d^{4.5}$-3$d^6$. The highest covalency is generally observed for the functionals employing correlation correction methods (i.e. PBE+$U$ and HSE), especially at higher 3$d$ count. The most highly active compounds for ORR[16,45,54] and OER[17,20,44] reside in the range of 3$d^{4.5}$-3$d^6$ (e.g. PBCO, BSCF2.625), and are consistent with being highly covalent compounds. This high degree of covalency may facilitate the high ORR and OER activities by multiple methods, but at least one contribution is potentially to enable high electrical conductivity in these compounds. The (O $p$ – TM $d$) band center values are referenced to the PBE (O $p$ – TM $d$) band center value to observe the effects of DFT functional in **Figure 2D**. PBEsol behaves nearly identically to PBE, and SCAN results in minor differences from PBE, with the TM $d$ (O $p$) band center being slightly above (below) the O $p$ (TM $d$) band center relative to PBE, with a crossing point at a 3$d$ electron count of about 4 (3$d^4$) (i.e. Mn$^{3+}$, Fe$^{4+}$). These differences in O $p$ and TM $d$ band center values are also present when comparing HSE with PBE, but for HSE are much larger than the comparison of SCAN and PBE. These larger band shifts of HSE than SCAN relative to PBE are consistent with differences in band shifts seen for the same set of functionals in work on $\beta$-MnO$_2$ and between BaTiO$_3$, PbTiO$_3$ and BiFeO$_3$.[30,52] Finally, PBE+$U$ results in different trends from that seen for SCAN and HSE, where the PBE+$U$ (O $p$ – TM $d$) band center values relative to PBE result in much higher (lower) O $p$ (TM $d$) band centers for the entire range of 3$d$ electron count. This trend is the result of the high O $p$-band centers for PBE+$U$ seen in **Figure 2A** and the low TM $d$-band centers (at least at high 3$d$ electron numbers) for PBE+$U$ in **Figure 2B**.

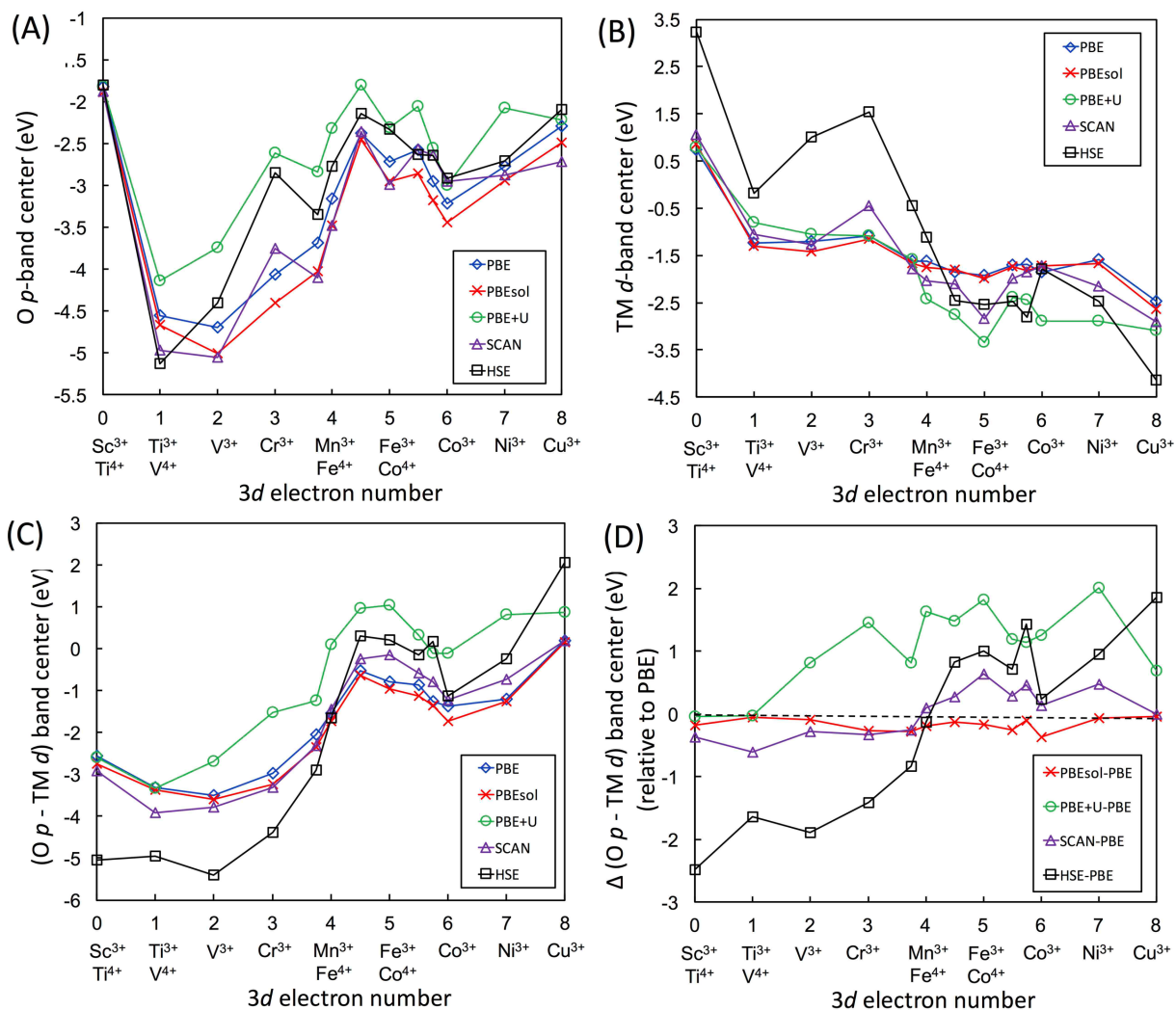

**Figure 2.** Plots of (A) O $p$-band center, (B) transition metal (TM) $d$-band center, (C) difference between O $p$-band center and TM $d$-band center (denoted here as (O $p$ – TM $d$) band center), and (D) difference between O $p$-band center and TM $d$-band center relative to PBE, all as a function of 3$d$ electron count of the transition metal. These 3$d$ electron values comprising the $x$-axis of were obtained assuming an ionic model where oxygen is in the $O^{2-}$ state and assigning formal valences to the cations. The different colored datasets correspond to values calculated using the different DFT functionals employed in this work. The solid lines connecting the data points are meant as guides for the eye, and the black dashed line in (D) marks a value of 0 for the change in respective band center difference relative to PBE. For all plots, the O $p$-band and TM $d$-band centers were calculated using all electronic states.

**2.2: Trends between calculated (DFT) and experimental (XES) O $p$-band centers**

X-ray emission and absorption spectroscopy methods have been successfully used to gain a deeper fundamental understanding of the interplay between the electronic structure and resulting OER catalytic activity of a series of perovskite materials. Specifically, the works of Suntivich *et al.*[53] and Hong *et al.*[38,39] have used these X-ray spectroscopic techniques to map the densities of occupied and unoccupied electronic states of numerous perovskites, and aligned these experimentally mapped densities of states to an absolute energy scale. With this information, one can directly map features of the perovskite electronic structure (e.g. Fermi edge location, band gap, bond hybridization, charge-transfer gap) to well-defined redox energy couples and the associated measured OER current density (at constant applied potential) or OER overpotential (at constant current density) to examine catalytic trends in a series of perovskites and draw connections between the catalytic activity of the perovskite and its electronic structure.

Here, we have examined the relationship of DFT-calculated O *p*-band centers and experimental O *p*-band centers obtained using XES data from the work of Hong *et al.*[39] These XES data have previously been shown to accurately represent the ground state densities of states,[38,55] as simulated through DFT, across the range of chemistries investigated here, and so can be correlated linearly to DFT calculations. **Figure 3A** through **Figure 3E** contain parity (predicted versus actual values) plots of DFT O *p*-band center versus XES O *p*-band center for each DFT functional, and contain DFT O *p*-band centers calculated in two ways: using all electronic states (blue circles) and only occupied electronic states (green triangles). Plotting the DFT O *p*-band centers obtained using all electronic states and only occupied electronic states was performed for two reasons: (1) XES experiments measure only the occupied electronic states, so it may be more reasonable to compare the XES O *p*-band center values with DFT O *p*-band center values obtained using only the occupied electronic states, and (2) it remains an open question whether the DFT O *p*-band center descriptor can generally yield better correlations with the physical quantities examined throughout this work when it is calculated using all electronic states or only the occupied electronic states.

From the trends and computed metrics listed in **Figure 3**, it is difficult to definitively claim which DFT functional results in the best correlations of O *p*-band center between theory and experiment. However, our analysis of trends between experimental XES and DFT-calculated O *p*-band centers demonstrates that while PBE, PBEsol and SCAN give comparable error and prediction metrics, PBE+$U$ and HSE perform markedly worse. PBEsol shows the predictive trend that is closest with experimental XES values as determined by an MAE value of the PBEsol data versus the XES data of 0.37 eV. SCAN shows the best linear correlation between calculated O $p$-band center and measured XES O $p$-band center with $R^2$ = 0.78, and PBE demonstrates a comparable linear correlation with $R^2$ = 0.75. In contrast, PBE+$U$ and HSE perform markedly worse. PBE+$U$ shows the largest errors relative to XES data, with a large MAE of 1.10 eV, which results from the large upshift of the PBE+$U$ O $p$-band centers in comparison to experimental data, which was discussed previously in **Section 2.1**. The work of Hong *et al.* also compared the DFT-calculated O $p$-band centers using PBE+$U$ methods with XES-measured O $p$-band centers. They found that PBE+$U$ tended to shift the O $p$-band center values up (i.e. less negative) relative to experiment by approximately 1 eV, consistent with the findings of this work. This large error is reduced somewhat when only the occupied O 2$p$ states are considered (compare the blue circle versus green triangle data in **Figure 3C**), however the resulting $R^2$ value is only 0.25. Perhaps surprisingly, HSE shows the worst agreement with experiment in terms of its $R^2$ value, with an $R^2$ = 0.31 and a MAE value relative to XES values of 0.75 eV. This is counterintuitive as the HSE functional contains the most physically accurate description of electron-electron interactions of the different functionals used in this study. This result suggests that HSE cannot be used as a parameter-free method to accurately reproduce at least some aspects of the electronic structure of perovskite oxides containing late transition metals. This observation is in agreement with previous studies which found that correctly reproducing the experimental bulk electronic structure of a series of perovskites with HSE can only be done by tuning the amount of Hartree-Fock exchange for each material.[56,57]

Based on the tabulated MAE values in **Figure 3F**, using only the occupied O 2$p$ states in the band center calculations tends to yield worse results. This result is unexpected, as XES only measures the occupied electronic states of the material, therefore one may expect that comparing the XES data with DFT-calculated band centers obtained by integrating only over occupied states should yield better correlations. It is not clear at this time why a better agreement is not obtained when using just occupied states, and further study of this issue would be of interest.

Overall, we have found that the use of PBE+$U$ and HSE, which have tunable parameters, are not generally reliable for obtaining accurate correlations between DFT-calculated and XES-measured O $p$-band centers. PBE, PBEsol and SCAN demonstrate lower errors than PBE+$U$ and HSE both in the linear fit and versus the XES values, as well as more robust linear correlations.

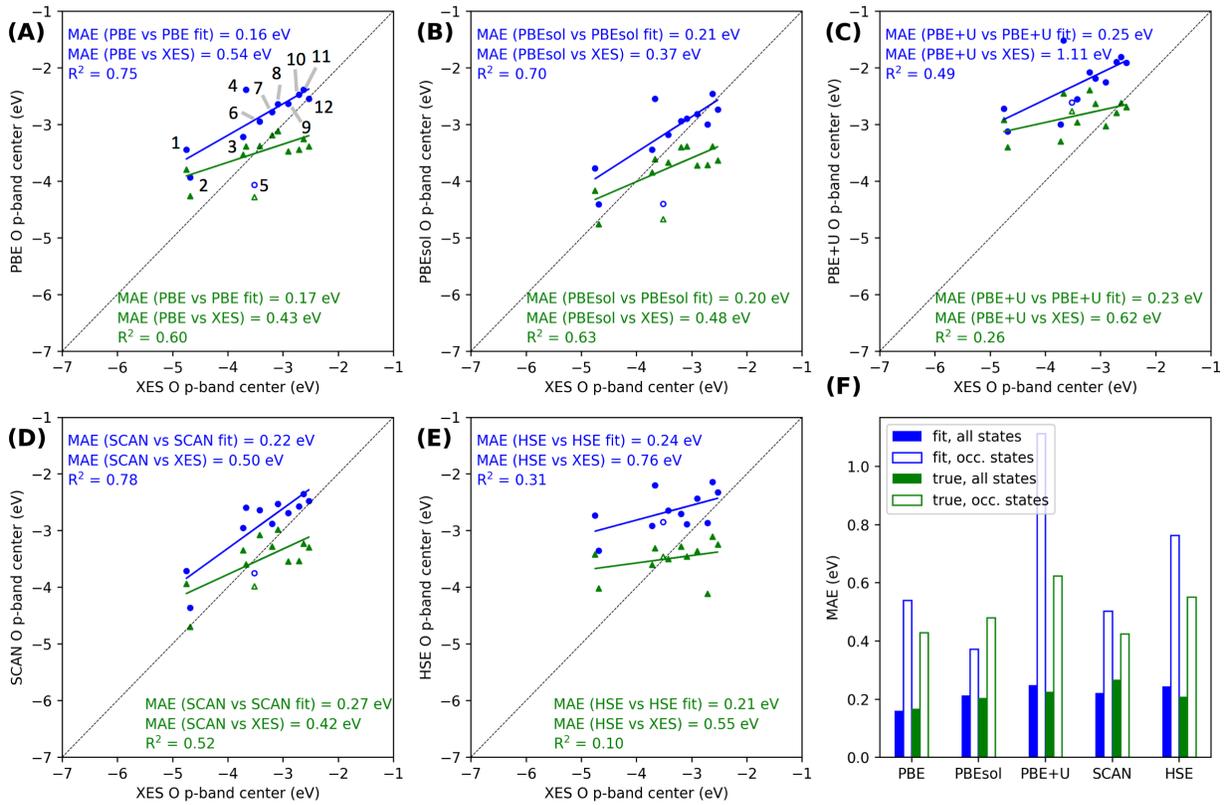

**Figure 3**. Parity plots of calculated bulk O *p*-band center for PBE (A), PBEsol (B), PBE+*U* (C), SCAN (D), and HSE (E) versus experimental XES O *p*-band center. The blue circle and green triangle data use DFT O *p*-band centers obtained by integrating over all electronic states and occupied electronic states only, respectively. The empty symbols denote LaCrO$_3$, which was not included in the linear fit due to suspected inaccuracy in the measured O *p*-band center value due to surface oxidation.[39] In (A-E), the $R^2$ and mean absolute error (MAE) values are reported, and the MAE values are tabulated in (F) for each functional. For each linear fit, there are two MAE values reported: one for the MAE in the linear fit (i.e. data points versus fitted line, e.g. "PBE vs. PBE fit") and another for the MAE between the DFT and experimental values (i.e. data points versus the *y* = *x* diagonal line, e.g. "PBE vs. XES"). These DFT vs. XES MAE values represent how far the calculated data points are away from the diagonal *y* = *x* line, which represents perfect agreement between theory and experiment. The material labels in (A) correspond to 1: LaFeO$_3$, 2: LaMnO$_3$, 3: LaCoO$_3$, 4: SrFeO$_{2.75}$, 5: LaCrO$_3$, 6: LSC25, 7: LaNiO$_3$, 8: LSC50, 9: SBCO, 10: GBCO, 11: BSCF2.625, 12: PBCO, and are in order of increasing XES O *p*-band center. All experimental XES data was obtained from the work of Hong *et al*.[39]

## 2.3: Correlation with high temperature oxygen surface exchange rates

The O *p*-band center has been successfully used as a descriptor for the surface exchange rate coefficient (*k**) of the high temperature ORR, and has enabled improved fundamental understanding of correlations between the bulk electronic structure and oxygen reduction catalytic activity of perovskites as well as materials discovery and design of new perovskite compounds for stable, high ORR activity compounds.[11,16,58] These previous studies used PBE+*U* methods to calculate the O *p*-band center used to predict *k** values, and here we compare the predictive ability of PBE+*U* against PBE, PBEsol, SCAN and HSE for obtaining experimental *k** values.[16]

PBE yields the best prediction of the experimental *k** values as a function of computed O *p*-band center based on the perovskites examined in this work, with an $R^2$ value of 0.87 and a MAE of 0.53 cm/s (in log units). PBE+*U* is slightly worse than PBE but also provides effective predictions, as shown in **Figure 4**. As with the comparison to experimental XES data in **Section 2.2** (**Figure 3**), HSE also shows the worst correlations (lowest $R^2$, highest MAE) with *k**. It should be mentioned that our previous work, including Lee *et al.*[11] and Jacobs *et al.*,[16] used PBE+*U* to calculate O *p*-band centers to predict *k** values, having $R^2$ values of 0.87 (Lee *et al.*, 9 different perovskites)[11] and 0.866 (Jacobs *et al.*, 19 different perovskites, 9 of which are the same as those in Lee, *et al.*),[16] which is higher than that of this work ($R^2$=0.85), most likely stemming from a different set and number of perovskites used in the linear fits. These results suggest that it might be advantageous to form correlations of *k** using PBE vs. PBE+*U* calculations of the O *p*-band center, as PBE is both slightly better in the present assessment and does not require fitting *U* parameter values for each transition metal, which yields different *U* values for different approaches.

Our assessments of the MAE values between predicted and experimental *k** values indicate that more accurate correlations are, just as in the case of the correlation of XES O *p*-band centers analyzed in **Section 2.2**, obtained when the O *p*-band center is calculated using all electronic states, and not when the O *p*-band center is calculated using only occupied electronic states (see **Figure 4F**). The better correlations of experimental *k** values with O *p*-band centers calculated using all electronic states result may be due to the additional electronic structure information related to the unoccupied states present in the O *p*-band center calculation using all states. This additional information is relevant because the process of oxygen surface exchange involves the transfer of oxygen and electrons to and from the perovskite, which will necessarily involve both bonding (occupied) and antibonding (unoccupied) oxygen electronic states. This process of charge transfer during oxygen exchange was explained in the context of a rigid band model in the work of Lee *et al.*,[11] in the form of molecular orbital hybridization arguments in Suntivich *et al.*[45] and using the magnitude of the charge-transfer energy in the work of Hong *et al.* The charge-transfer energy in the work of Hong *et al.* is directly related to the average energy of the occupied and unoccupied O 2$p$ states.[39]

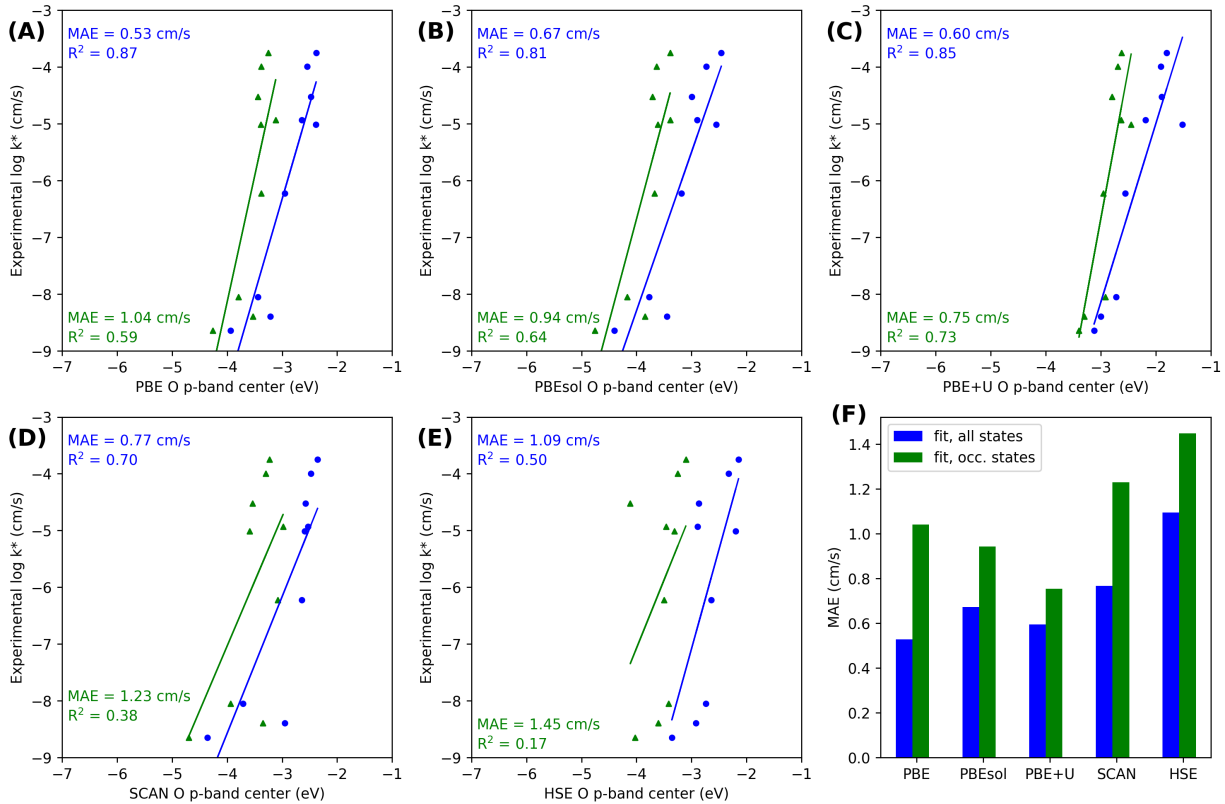

**Figure 4.** Plots of experimental high temperature oxygen exchange coefficient $k^*$ versus calculated bulk O $p$-band center for PBE (A), PBEsol (B), PBE+$U$ (C), SCAN (D), and HSE (E). The blue circle and green triangle data use DFT O $p$-band centers obtained by integrating over all electronic states and occupied electronic states only, respectively. In (A-E), the $R^2$ and mean absolute error (MAE) values are reported, and the MAE values are tabulated in (F) for each functional. The material labels in (A) correspond to 1: $LaMnO_3$, 2: $LaCoO_3$, 3: $LaFeO_3$, 4: LSC25, 5: $SrFeO_{2.75}$, 6: LSC50, 7: GBCO, 8: PBCO, 9: BSCF2.625, and are in order of increasing $k^*$ value. The experimental $k^*$ values were all obtained from the work of Jacobs *et al.*, and correspond to operating conditions of approximately $T = 1000$ K and $p(O_2) = 0.2$ atm.[16]

## 2.4: Correlation with aqueous oxygen evolution reaction current densities

In addition to modeling the high temperature ORR surface exchange rates, the O $p$-band center has also been used to form correlations of perovskite catalytic activity for the aqueous OER.[17,20] The work of Grimaud *et al.* demonstrated that the OER overpotential (at constant current density) varied linearly with O $p$-band center calculated with GGA+$U$ methods for a series of Co-based perovskites.[20] Here, we demonstrate the ability of the O $p$-band center to predict the current density of the OER measured at constant potential (in this case, 1.6 V versus the Reference Hydrogen Electrode (RHE)), and again compare the predictive ability of PBE, PBEsol, PBE+$U$, SCAN and HSE.

From **Figure 5**, it is clear that PBE produces the best correlation between OER current density and the O $p$-band center, with an $R^2$ value of 0.81 and MAE of 0.24 mA/cm$^2$ (in log units). It is also noteworthy that, as with the XES data in **Figure 3** (**Section 2.2**) and the ORR $k^*$ values in **Figure 4** (**Section 2.3**), that HSE once again demonstrates the worst correlation between the experimental quantity of interest and the calculated O $p$-band center, with an $R^2$ value of 0.29 and MAE of 0.44 mA/cm$^2$. In addition, the tabulated MAE values in **Figure 5F** again show that using all electronic states in the calculation of the O $p$-band center result in lower MAE values for all functionals. This result of the use of all electronic states in the band center calculation resulting in lower MAE values is consistent with all electronic states resulting in lower MAEs (than using occupied states only) for the XES and $k^*$ correlations discussed in **Section 2.2** and **Section 2.3**, respectively. We believe the improved correlations of OER current density with O $p$-band center values calculated using all electronic states can be explained by the same mechanism we discussed in **Section 2.3** in the context of improved correlations with $k^*$ using all electronic states. Specifically, the OER and ORR processes both involve the transfer of oxygen and electrons to and from the perovskite, which necessarily involves unoccupied oxygen states.

The previous work of Grimaud *et al.* used GGA+$U$ calculations of the O $p$-band center to assess correlations in the OER overpotential for a series of Co-based perovskites.[20] They demonstrated a linear correlation exists, and while no $R^2$ value was explicitly reported, from extracting their reported data we calculated an $R^2$ value of approximately 0.96. We note the metrics used to assess the catalytic activity of perovskites for the OER differ between this work (OER current density at constant voltage, data from Hong *et al.*)[39] and the work of Grimaud *et al.* (OER potential at constant current density),[20] however qualitative comparisons can still be made. The $R^2$ using GGA+*U* O *p*-band centers in Grimaud *et al.* is much higher than the PBE+*U* $R^2$ of 0.61 shown in **Figure 5C**. While some minor differences may arise by examining correlations using data based on constant voltage versus constant current and the use of different pseudopotentials in the work of Grimaud *et al.* (PW-91) versus this work (PBE), we believe the major difference arises from the range of perovskite compositions used to construct the correlation. The work of Grimaud *et al.* clearly shows that a good linear correlation can be obtained using GGA+*U* for the narrow chemical space involving only Co-based perovskites with 3*d* electron numbers between $3d^{5.5}$ (e.g. PBCO) and $3d^6$ (LaCoO$_3$). However, in this work we have used a broader chemical range of compounds with 3*d* electron numbers ranging from $3d^3$ (LaCrO$_3$) to $3d^7$ (LaNiO$_3$). Therefore, it is apparent that the correlations obtained using PBE+*U* become significantly worse when considering a chemically diverse set of perovskite materials that span a wide range of 3*d* electron count. In addition, Man *et al.* have calculated the theoretical OER potential using revised PBE (RPBE) methods and compared the values to experimental OER potentials, and have demonstrated a strong linear trend of experimental versus theoretical OER potential.[21] Lee *et al.* have calculated the theoretical OER potential using PBE+*U* methods and also made comparisons to experimental OER potentials and to the work of Man *et al.*[19] While no $R^2$ values for the plots of experimental versus theoretical OER potential in the work of Lee *et al.* were given, extracting their data and using a linear fit produced $R^2$ values of 0.91 and 0.80 for the RPBE data and PBE+*U* data, respectively, where we obtain these $R^2$ values by only plotting the BO$_2$-terminated data shown in the work of Lee *et al.* for consistency.[19] Overall, it is apparent that to obtain the best correlative trend of OER activity using the O *p*-band center (calculated using all electronic states) for a broad range of perovskite compounds, PBE emerges as the method of choice.

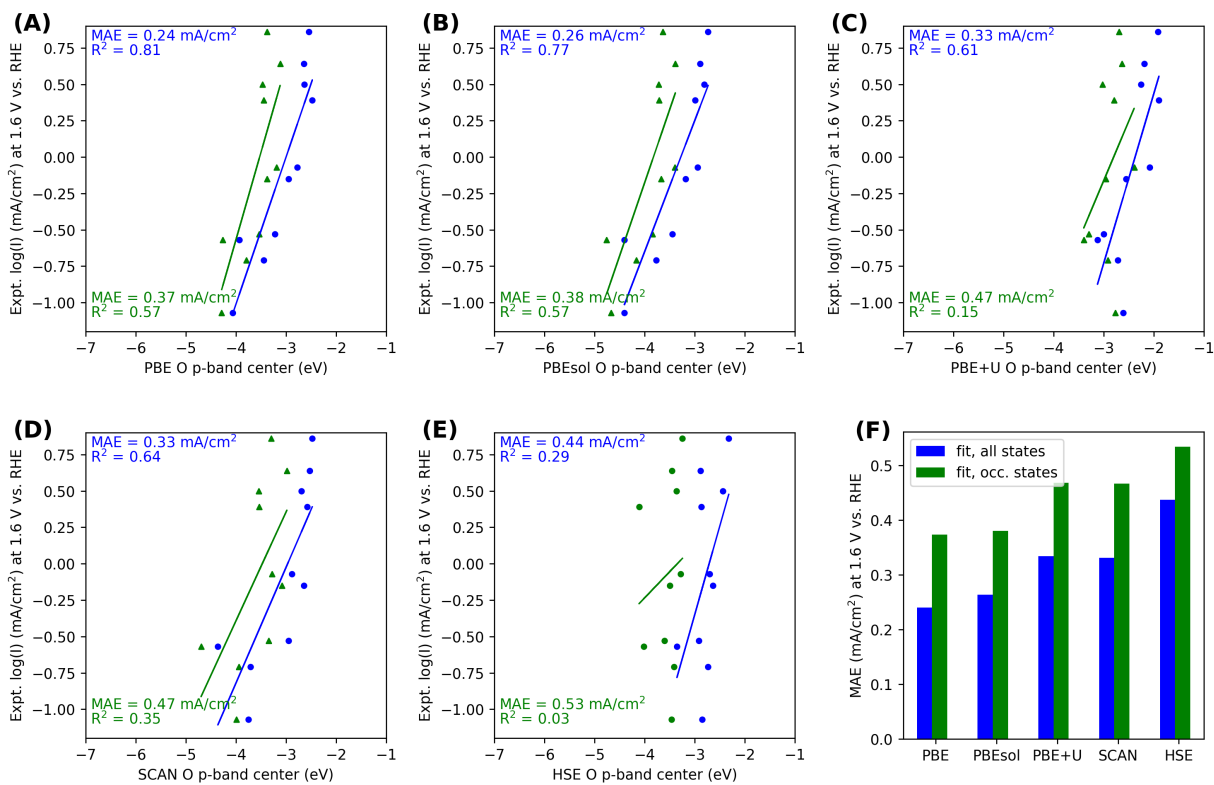

**Figure 5.** Plots of the logarithm of experimental aqueous OER current density versus calculated bulk O *p*-band center for PBE (A), PBEsol (B), PBE+$U$ (C), SCAN (D), and HSE (E). The blue circle and green triangle data use DFT O *p*-band centers obtained by integrating over all electronic states and occupied electronic states only, respectively. In general, a perovskite exhibiting a higher current density at constant potential is indicative of a more active catalyst. In (A-E), the $R^2$ and mean absolute error (MAE) values are reported, and the MAE values are tabulated in (F) for each functional. All of the experimental current density data were obtained at a constant voltage of 1.6 V versus the reference hydrogen electrode (RHE). The material labels in (A) correspond to 1: $LaCrO_3$, 2: $LaFeO_3$, 3: $LaMnO_3$, 4: $LaCoO_3$, 5: LSC25, 6: $LaNiO_3$, 7: GBCO, 8: SBCO, 9: LSC50, 10: PBCO, and are in order of increasing OER current density value. All experimental OER current density data was obtained from the work of Hong *et al*.[39] The average error in the experimental OER current density is 0.14 mA/cm$^2$ (in log units).

**2.5: Correlations of OER intermediate binding energies**

In addition to the correlation of OER current density (at constant potential) with the overall catalytic rate of a material, substantial work has been done to correlate the binding energies of intermediate species formed during the OER with the overall OER activity.[18,19,21–23,59,60] However, it is unclear whether robust correlations exist between intermediate binding energies and the DFT-calculated bulk O *p*-band center. In this section, we examine the trend of the binding energies of OER intermediate species as a function of O *p*-band center using data available from previous studies. We have integrated data on the binding energies of *O and *OH (here * indicates a surface bonding site for the adsorbed species) on the (001) $BO_2$-terminated surface of a series of perovskite materials. These studies include: Man *et al.*,[21] who used RPBE techniques of La(Cr, Mn, Fe, Co, Ni, Cu)$O_3$; Santos *et al.*[22] used both PBE and HSE methods for La(Cr, Mn, Fe, Co, Ni)$O_3$; Wang *et al.*[23] did a comparative analysis using PBE, PBE+*U*, and HSE on La(Cr, Mn, Fe)$O_3$; and Lee *et al.*,[19] who used PBE+*U* methods to compare binding energies of $BO_2$ and AO-terminated surfaces of La(Cr, Mn, Fe, Co, Ni)$O_3$. **Figure 6** contains plots of all of the binding energy data collected from these studies, grouped by DFT functional used: PBE (**Figure 6A**), PBE+*U* (**Figure 6B**) and HSE (**Figure 6C**). We note that Man *et al.*[21] used RPBE and Santos *et al.*[22] and Wang *et al.*[23] used PBE, however the agreement in the binding energies between these studies is very good. In addition, we assume that the O *p*-band center values for PBE apply to the RPBE binding energy data, though we acknowledge that there is some minor uncertainty in the positions of the O *p*-band centers for the data from Man *et al.*, probably on the order of 100 meV based on previous comparisons of O *p*-band centers calculated using different GGA-based methods.[11,16]

Our analysis of trends in intermediate binding energies as a function of the O *p*-band center for PBE, PBE+*U* and HSE indicate that strong linear correlations exist when correlating binding energies using PBE, while no such trend appears to exist for PBE+*U* and HSE, at least given the current available intermediate binding energy data. From **Figure 6A**, PBE results in strong linear correlations of *O ($R^2$ = 0.87), *OH ($R^2$ = 0.87), and the difference *O-*OH ($R^2$ = 0.79) binding energies with the O *p*-band center. We note here that while binding energies calculated in the work of Santos *et al.*,[22] Wang *et al.*[23] and Man *et al.*[21] generally agree with each other within a couple hundred meV, the calculated data in the work of Montoya *et al.*[18] differs quantitatively (on the scale of ~ 1 eV) from these other studies, which could be due to their use of the cubic perovskite structure (with no internal octahedral tilting), resulting in larger binding energy differences. By comparison with PBE, no such linear correlation is apparent for the PBE+*U* (**Figure 6B**) and HSE (**Figure 6C**) cases. One reason for the lack of correlation in the PBE+*U* and HSE cases may be due to the lesser availability of binding energy data using these functionals. However, from observing the sequence of the TM elements in the perovskites when moving from left to right in the direction of increasing O *p*-band center in **Figure 6B** and **Figure 6C**, the sequence of the TM elements in the perovskites differs from the PBE results of **Figure 6A**. In contrast, the sequence of TM elements in the perovskites as a function of O *p*-band center in **Figure 6A** directly follows the progression from left to right along the 3*d* row of the periodic table. The observed progression of increasing O *p*-band center when moving from left to right along the 3*d* row observed for the PBE results in **Figure 6A** is largely in agreement with the progression of experimentally measured O *p*-band centers from XES (see **Section 2.2** and **Figure 3)**. That is, for the PBE data in **Figure 6A** and from the XES data in **Figure 3**, when progressing from low to high O *p*-band center the progression of TM elements in the perovskites is Mn ≈ Fe < Co < Ni. The observed inability of PBE+*U* and HSE to accurately capture the trend of O *p*-band center for a series of TM elements results in intermediate binding energy data that appears largely flat or with a slight negative slope as a function of O *p*-band center, in clear contrast to the strong positive slope of the PBE data in **Figure 6A**. That is, for the PBE+*U* and HSE data in **Figure 6B** (**Figure 6C**), when progressing from low to high O *p*-band center the progression of TM elements in the perovskites is Mn < Co < Fe < Ni. Further support for the accuracy of the PBE results is given by the fact that the order of material binding energies qualitatively agrees with experimental results of OER overpotential and their correlation with calculated trends of intermediate binding strength for these compounds.[19–21,44,59] More specifically, in the work of Man *et al.*,[21] the trend in OER overpotential indicates that OER activity for different TM elements in lanthanide perovskites increases in the order of Mn < Fe < Co < Ni. This trend is the same as the sequence of TM elements seen for the PBE results in **Figure 6A**, and does not agree with the PBE+*U* and HSE results in **Figure 6B** and **Figure 6C**, respectively. Therefore, it is again the case that PBE has demonstrated the best correlations with O *p*-band center of the different functionals and available experimental data analyzed here, though we note that some improvement in the PBE+*U* and HSE results is likely if data of more systems would be included.

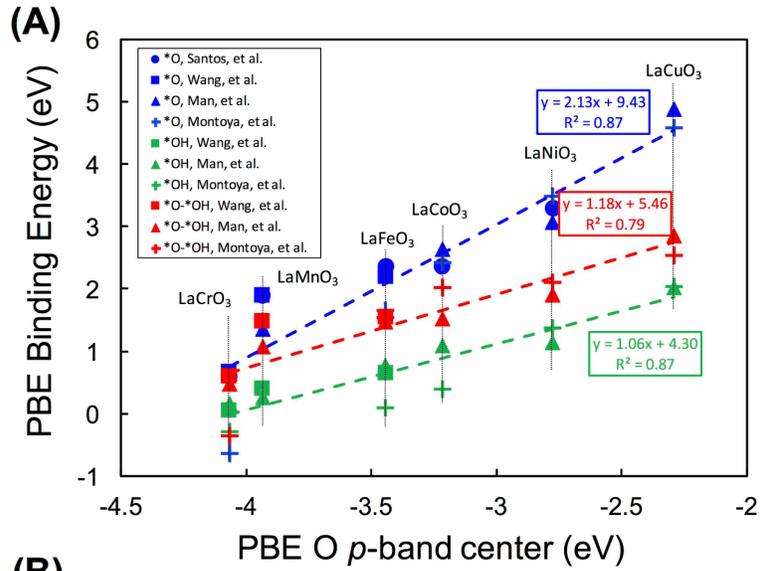
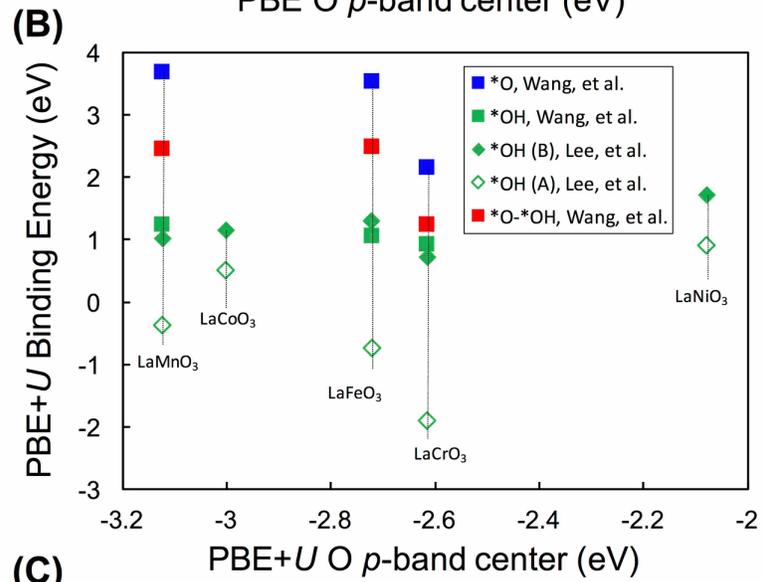
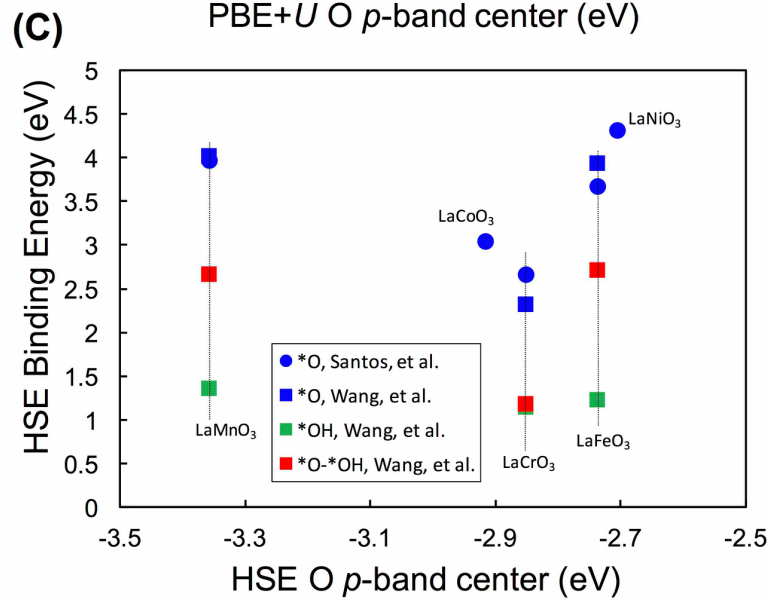

**Figure 6.** Binding energy of OER intermediates *O, *OH, and the difference *O - *OH of assorted perovskites as a function of bulk O $p$-band center for (A) PBE (B) PBE+$U$ and (C) HSE. In each figure, the blue, red, and green points denote binding energies for *O, *O - *OH, and *OH, respectively. The circle, square, triangle, diamond and cross symbols denote data obtained from the work of Santos et al.,[22] Wang et al.,[23] Man et al.,[21] Lee et al.[19] and Montoya et al.[18], respectively. The binding energies were all obtained for (001) $BO_2$ perovskite surfaces, except in (B), where the data from Lee et al. encompassed both $BO_2$ (labeled as (B), solid symbols) and AO (labeled as (A), empty symbols) surfaces.

## 2.6: Additional discussion

Throughout this work, we have used the O $p$-band center calculated using the bulk density of states. However, from **Figure 6B**, it is clear that the bulk O $p$-band center fails to differentiate the *O and *OH binding energies of different surface terminations. In **Figure 6B**, the binding energy data of the (001) $BO_2$ surface (solid symbols) tend be 1-3 eV more positive (indicating weaker binding) than the (001) AO surface (empty symbols), depending on the material under consideration. Lee et al.[19] instead used the O $p$-band center of the subsurface layer of O atoms, and demonstrated that the *O and *OH binding energies of both surface terminations fall on a single linear trend line. Therefore, in considering catalytic quantities that are impacted by contributions from changing surfaces within a dataset, the use of bulk descriptors is likely to fail quantitatively (though qualitative insight might still be obtained), and the use of surface-sensitive descriptors is likely needed for the creation of quantitative correlations.

Given the importance for surfaces for many catalytic properties, and the inability of a bulk descriptor to distinguish behavior of different surfaces for the same bulk, it is interesting that quantities such as $k^*$ and the OER current density can be well-described by the bulk (PBE) O $p$-band center. We speculate that these quantities can be accurately described by bulk descriptors because the measured $k^*$ and the OER current density may be representative of catalysis occurring on corresponding surfaces (or sites) for all the materials. This may be reasonable as $k^*$ and the OER current density both represent rates of activated processes, that is, their values depend exponentially on the energetics of the oxygen reduction ($k^*$) and oxygen evolution (OER current density) reactions, respectively, which means that their values are likely dominated by the surface with the lowest energy barriers for the ORR and OER. If it is true that, to a first approximation, a single surface/set of sites is responsible for all of the catalysis occurring across a set of materials, then the measured value of $k^*$ or OER current density will be well-described by the bulk properties, assuming that the shift of the O *p*-band center from bulk to the catalytically-active surface/set of sites of interest is approximately the same for all materials considered. This approximately constant shift of the O *p*-band center from bulk to surface may also be the reason why the intermediate binding energies of the (001) $BO_2$ surface of numerous perovskites trend well with the bulk O *p*-band center (**Figure 6**). As a further example of how bulk and surface behavior correlate, it has been shown that the bulk O *p*-band center, the surface O *p*-band center, and adsorbate binding energies (*O, *OH, *OOH) scale well with each other.[11,19] In addition, the work of Calle-Vallejo *et al.* demonstrated the scaling of adsorbate binding energy with bulk oxide formation energy in transition metal oxides, further supporting this bulk-surface property relation.[61] In perovskite oxides, the work of Hong *et al.* found that surface hydroxylation trends across perovskite chemistries can be rationalized by the position of the bulk oxide Fermi level in relation to the OER potential. Lowering of the Fermi level (raising of the O *p*-band center) compared to the OER potential results in weaker hydroxide affinity during OER due to the ability of low charge-transfer gap oxides to screen charge, reducing the need to compensate charge with hydroxide ions.[39] The findings of these previous studies rationalize the correlations of bulk electronic structure, such as bulk O *p*-band center and covalency, with surface electrocatalytic properties and activities. These findings offer qualitative physical insight of catalyst properties which enables their practical use for catalyst design.

Given the above arguments that bulk descriptors can capture surface-sensitive catalytic trends, if a new material had a *k** or OER current density value that differed markedly from the established trend, this would be an indication that the ORR or OER process for that material may be occurring via a different mechanism (such as a different catalytically-active surface) than the materials comprising the trend. For example, in the work of Hong *et al.* it was found that the charge transfer energy can be used to discern which materials may facilitate the OER via different mechanisms, where it was found that $LaCrO_3$ was electron-transfer limited and PBCO was proton-transfer limited.[39] Another example of where discrepancies in correlations can give insight into catalytic mechanisms was performed by Jacobs *et al.*,[16] who showed that many *p*-type perovskites have *k** values that do not follow the same trend as *k** for *n*-type materials as a function of O *p*-band center, consistent with their operating by a different mechanism that involves at least some significant electron transfer limitation.

Overall, it is clear that while the bulk electronic structure correlations established in this work are useful for qualitative insight of catalytic properties and perovskite catalyst design, there are still numerous open questions related to establishing quantitative electronic structure trends with different catalytic processes for a range of perovskite chemistries. Some of these areas of further research may include understanding the coupling between the bulk and surface electronic structure in greater detail, and the evolution of the surface and electronic structure under environmental conditions relevant for catalysis. Some of these catalysis-related operational environments may include: an electrolytic cell in aqueous solution, an SOFC operating at high temperature, or the influence of cathodic bias. Experimental approaches employing *in situ* techniques to investigate perovskite oxide catalysts have enabled the monitoring of surface A-site enrichment under SOFC-relevant conditions with ambient-pressure X-ray photoelectron spectroscopy (AP-XPS),[62] and have shown evidence of lattice oxygen involvement in aqueous OER with on-line electrochemical mass spectroscopy (OLEMS).[63] *In operando* crystal truncation rod (CTR) measurements on oriented $RuO_2$ OER catalysts have provided evidence for previously unobserved reaction intermediates, such as hydroxyl-stabilized *OO species,[64] indicating that mechanistic understanding will play a central role in understanding how bulk electronic structure can influence surface processes in oxide catalysts. Further fundamental understanding of the interplay of these phenomena with the underlying perovskite electronic structure and trends in catalytic properties would further enhance the quantitative accuracy and utility of electronic structure descriptors for perovskite-based catalysis and electrocatalysis.

Finally, we would like to remark on the general pattern uncovered throughout this work, in which we found PBE results in the best correlations of the different catalytic measures with the calculated O *p*-band center. As HSE and PBE+*U* have been historically demonstrated to be more accurate in capturing a wide range of physics in transition metal oxide compounds (such as formation energies,[65–68] redox energies,[47–51] and bandgaps[69–72]), it is surprising that PBE shows the best correlations with various catalytic measures for these materials. It is not our intention to claim that PBE captures catalysis-relevant physics more accurately than PBE+*U* or HSE (it likely does not), but only that PBE produces better correlations between O *p*-band determined by first-principles theory and multiple experimental measures of catalytic ability. We speculate that the approximate corrections associated with PBE+*U* and HSE, while helping for many physical quantities, may create significant errors in the relative band level energies, and thus result in reduced correlation with the measured catalytic properties examined throughout this work. Part of the reason we observe decreased correlative ability of PBE+$U$ and HSE may be related to our current focus on more covalent compounds (e.g. perovskites with TM=Cr-Ni), which are predominantly metallic perovskites and may not be treated as well as insulators with the PBE+$U$ and HSE approaches. In addition, there may be physics at the surface not captured well by bulk-fitted $U$ values and exchange fractions that lead to additional errors in the correlations established with these methods.

## 3. Summary and Conclusions

This study evaluated the effectiveness of the O $p$-band center as a bulk electronic structure descriptor to form correlations for numerous measures of perovskite catalytic activity. The effectiveness of the O $p$-band center was examined across five different DFT functionals: PBE, PBEsol, PBE+$U$, SCAN and HSE for a set of up to 20 technologically relevant perovskite oxides. We have assessed trends between the DFT-calculated O $p$-band center with the experimental XES O $p$-band center and numerous measures of perovskite catalytic activity, including: the high temperature ORR surface exchange rate, the aqueous OER current density (at constant applied potential), and the binding energies of OER intermediates. We have also examined trends of O $p$-band and TM $d$-band center as a function of $3d$ electron number and DFT functional.

Surprisingly, we found that the best correlative trends of the different measures of catalytic activity with the calculated O $p$-band center emerged when PBE was used, and PBE was thus found to be better than PBEsol, PBE+$U$, SCAN and HSE at forming these correlations. Second, we have found that PBE successfully yields strong linear correlations between the O $p$-band center and every measure of perovskite catalytic activity examined here, with correlations of the PBE data yielding the high $R^2$ ($R^2 > 0.8$) and low MAE values for all catalytic measures. Therefore, the modifications to the electronic structure of transition metal-containing perovskite oxides due to Hubbard $U$ (for PBE+$U$) and Hartree-Fock exchange fraction (for HSE) do not generally yield improved correlations and predictions when using the O $p$-band descriptor, at least for the $U$ and exchange fraction values used in this work. Given that previous studies have successfully employed PBE+$U$ on different subsets of perovskites,[11,16,19,20] we remark here that while reasonable trends were observed with our PBE+$U$ data, we believe that the use of PBE will result in more reliable trends when considering sets of perovskites spanning a wide composition or chemical range, and will aid in transferability of results between studies as no fitted $U$ values are required.

Overall, this work has demonstrated that strong linear correlations between the bulk O $p$-band center, calculated using standard DFT-PBE calculations, can be made for a range of measures of perovskite catalytic activity for both oxygen reduction and evolution reactions under a single consistent computational framework. This work therefore provides useful guidance for future experimental assessment of perovskite catalytic activity and correlations, as well as a set of computational best-practices to yield the most accurate correlations of perovskite catalytic activity using the O $p$-band center as an electronic structure descriptor, therefore aiding in the future design and understanding of perovskite oxides for use in electrochemical energy applications.

## 4. Computational methods

All Density Functional Theory (DFT) calculations were performed using the Vienna Ab-Initio Simulation Package (VASP) code.[73] Exchange and correlation functionals of Perdew-Burke-Ernzerhof (PBE),[24] PBE parameterized for solids (PBEsol),[25] PBE with Hubbard $U$ correction (PBE+$U$),[26–28] the Strongly Constrained and Appropriately Normed functional (SCAN),[29,30] and the hybrid functional of Heyd-Scuseria-Ernzerhof (HSE06)[31,32] were used. The $U$ values in PBE+$U$ were the same as those used in the Materials Project: $U$ = 3.25, 3.7, 3.9, 5.3, 3.32, and 6.7 eV for V, Cr, Mn, Fe, Co, and Ni, respectively.[33] HSE calculations were performed using a cutoff of 0.2 Å$^{-1}$ and Hartree-Fock exchange fraction of 25% (standard HSE06). All materials were first fully relaxed (volume and ions) in their conventional cell form. After this initial full relaxation, relaxed structures were used to construct 2×2×2 supercells consisting of 5 perovskite formula units (40 atoms/cell), and the lattice parameters were changed to pseudocubic (pc), where $a_{pc} = b_{pc} = c_{pc} = (V_{full\ relax})^{1/3}$, and a subsequent internal relaxation of ions only was performed. All materials were simulated as ferromagnetic to have a consistent set of calculations across many chemistries. The structural consistency from using pseudocubic simulation cells with ferromagnetic ordering allows us to effectively examine and understand electronic structure trends both as a function of composition and DFT functional used. Integration in the Brillouin zone was performed using a Monkhorst-Pack $k$-point mesh of 4×4×4 for all 2×2×2 40-atom pseudocubic cells.[74] For full relaxations of conventional unit cells, Monkhorst-Pack $k$-point meshes of 12×12×12, 8×8×6 and 8×8×8 were used for 5-atom cubic (spacegroup $Pm\overline{3}m$), 20-atom orthorhombic (spacegroup $Pbnm$), and 10-atom rhombohedral (spacegroup $R\overline{3}c$) cells, respectively. For all calculations, projector augmented wave (PAW)[75] Perdew-Burke-Ernzerhof (PAW-PBE) type pseudopotentials matching those used in the Materials Project database were used, and the planewave cutoff energy was set to 500 eV. All key VASP input and output files, and all data for all figures, can be found as part of the **SI** and are also publicly available on Figshare (see **Supporting Information** summary section for Figshare link).

All initial perovskite structures were obtained from the Materials Project database,[33] except for the doped compounds, which were not available on the Materials Project database at the time of this writing. LSM25 was made using $LaMnO_3$ as a starting structure. LSC25, LSC50, and BSCF2.625 were made using $LaCoO_3$ as a starting structure. PBCO, GBCO and SBCO were made also using $LaCoO_3$ as a starting structure, and the rare earth and Ba atoms were ordered into layers to form the double perovskite structure. For all materials containing O vacancies, the vacancies were created in the AO row following previous studies.[11,20]

The O $p$-band center is the centroid of the oxygen $2p$ electronic states, given relative to the Fermi level. The O $p$-band center values were calculated following previous studies by integrating over the oxygen $2p$ projected density of states (PDOS).[11,15,16] The calculation of the O $p$-band center using all electronic states was performed by integrating from negative to positive infinity in energy of the PDOS. The calculation of the O $p$-band center using occupied states only was performed by integrating from negative infinity up to the Fermi energy of the PDOS. The transition metal $d$-band center is the centroid of the transition metal $3d$ electronic states, also given relative to the Fermi level. It was calculated in an analogous way as the O $p$-band center. For materials containing more than one transition metal element, the overall $d$-band center was calculated as the composition-weighted average of the $d$-band centers of each transition metal element in the compound.

## Acknowledgements

R. Jacobs and D. Morgan were supported by the National Science Foundation Software Infrastructure for Sustained Innovation (SI2) award #1148011 and J. Hwang was supported by Eni. This research used computing resources of the National Energy Research Scientific Computing Center (NERSC), which is supported by the U.S. Department of Energy Office of Science. This research was performed using the compute resources and assistance of the University of Wisconsin-Madison Center For High Throughput Computing (CHTC) in the Department of Computer Sciences.

## Conflicts of Interest

The authors declare no competing financial interest.

## Supplementary Information

Supplementary information is available detailing the calculated densities of states for all materials and all DFT functionals considered in this work, data on trends in the O *p*-band center for each DFT functional as well as key VASP input/output calculation files for all materials examined in this work. An excel file is also provided which contains the data used to create the figures shown in this work. This Excel file and all key data and VASP calculation files are publicly available on Figshare (https://figshare.com/s/15dde55390bd072bbacc).

## References

(1)  Hammer, B.; Norskov, J. K. Why Gold Is the Noblest of All the Metals. *Nature* **1995**, *376* (6537), 238–240.

(2)  Hammer, B.; Nørskov, J. K. Electronic Factors Determining the Reactivity of Metal Surfaces. *Surf. Sci.* **1995**, *343* (3), 211–220.

(3)  Hammer, B.; Norskov, J. K. Theoretical Surface Science and Catalysis — Calculations and Concepts. *Adv. Catal.* **2000**, *45*, 71–129.

(4)  Ruban, A.; Hammer, B.; Stoltze, P.; Skriver, H. L.; Nørskov, J. K. Surface Electronic Structure and Reactivity of Transition and Noble Metals. *J. Mol. Catal. A Chem.* **1997**, *115* (3), 421–429.

(5)  Greeley, J.; Jaramillo, T. F.; Bonde, J.; Chorkendorff, I.; Schlapbach, J. K. N. Computational High-Throughput Screening of Electrocatalytic Materials for Hydrogen